\documentclass[aps,prl,twocolumn,showpacs,reprint]{revtex4-1}
\usepackage{bm,bbm,amsmath,amssymb,amsbsy,graphicx,amsthm,color,amsthm,hyperref,txfonts}
\usepackage[english]{babel}
\providecommand{\ket}[1]{|#1\rangle}
\providecommand{\bra}[1]{\langle#1|}
\providecommand{\kebra}[2]{|#1\rangle\langle#2|}
\providecommand{\proj}[1]{|#1\rangle\langle#1|}
\begin{document}
\title{Coherently opening a high-Q cavity}
\author{Tommaso Tufarelli$^1$, Alessandro Ferraro$^2$, Alessio Serafini$^3$, Sougato Bose$^3$ and M. S. Kim$^1$} 
\affiliation{$^1$ QOLS, Blackett Laboratory, Imperial College London, SW7 2BW, UK; \\
$^2$ School of Mathematics and Physics, Queen's University Belfast, Belfast, BT7 1NN, UK; \\
$^3$ Department of Physics and Astronomy, University College London, Gower Street, London WC1E 6BT, UK.}

\begin{abstract}
We propose a general framework to effectively `open' a high-Q resonator, that is, to release the quantum state initially prepared in it in the form of a traveling electromagnetic wave. This is achieved by employing a mediating mode that scatters coherently the radiation from the resonator into a one-dimensional continuum of modes such as a waveguide. The same mechanism may be used to `feed' a desired quantum field to an initially empty cavity. Switching between an `open' and `closed' resonator may then be obtained by controlling either the detuning of the scatterer or the amount of time it spends in the resonator. First, we introduce the model in its general form, identifying (i) the traveling mode that optimally retains the full quantum information of the resonator field and (ii) a suitable figure of merit that we study analytically in terms of the system parameters. Then, we discuss two feasible implementations based on ensembles of two-level atoms interacting with cavity fields. In addition, we discuss how to integrate traditional cavity QED in our proposal using three-level atoms.
\end{abstract}
\maketitle
\noindent\textit{Introduction} --- The past two decades have witnessed the blooming of cavity-QED, through vast advances in the development of high-Q optical and microwave cavities, and in the ability to coherently control individual quantum emitters interacting with confined radiation \cite{q-optics,haroche,kimble}. Cavity-QED has long been the paradigmatic setup to investigate models of interaction between light and matter at the single-photon level, and led both to investigations into the fundamental properties of quantized radiation \citep{HarocheBook} and to the development of some of the most sophisticated quantum control techniques available to date \cite{sayrin}. Recently, analogous models have been implemented in a variety of experimental platforms such as circuit-QED, trapped ions and Bose-Einstein condensates (BECs) \cite{circuit-rev,ions-review,BECs}. 

\noindent Promising as this progress may be, the step from proof of principle demonstrations to operational quantum technology inspired by the cavity QED paradigm is prevented at this stage by a fundamental difficulty: on one hand, such an endeavor would request one to operate light matter interactions in the strong coupling regime, where the coupling strength is at least comparable to the cavity decay rate; on the other hand, it would be highly desirable to extract the state of the cavity field on reasonably short time scales. These two requirements, implying respectively high and low Q-factors, are inherently contradictory. In addition, in high-Q cavities, the photon lifetime can be maximized only by reducing the transmittivity to a minimum, typically to values comparable to the cavity losses. Thus, the physically accessible field that naturally leaks out from the cavity does not faithfully retain the quantum properties of the intra-cavity field, posing a major problem for the exploitation of cavity QED-like architectures in scalable quantum information processing and quantum networks \cite{kimble08}.
\begin{figure}[t!]
\begin{center}\includegraphics[width=.75\linewidth]{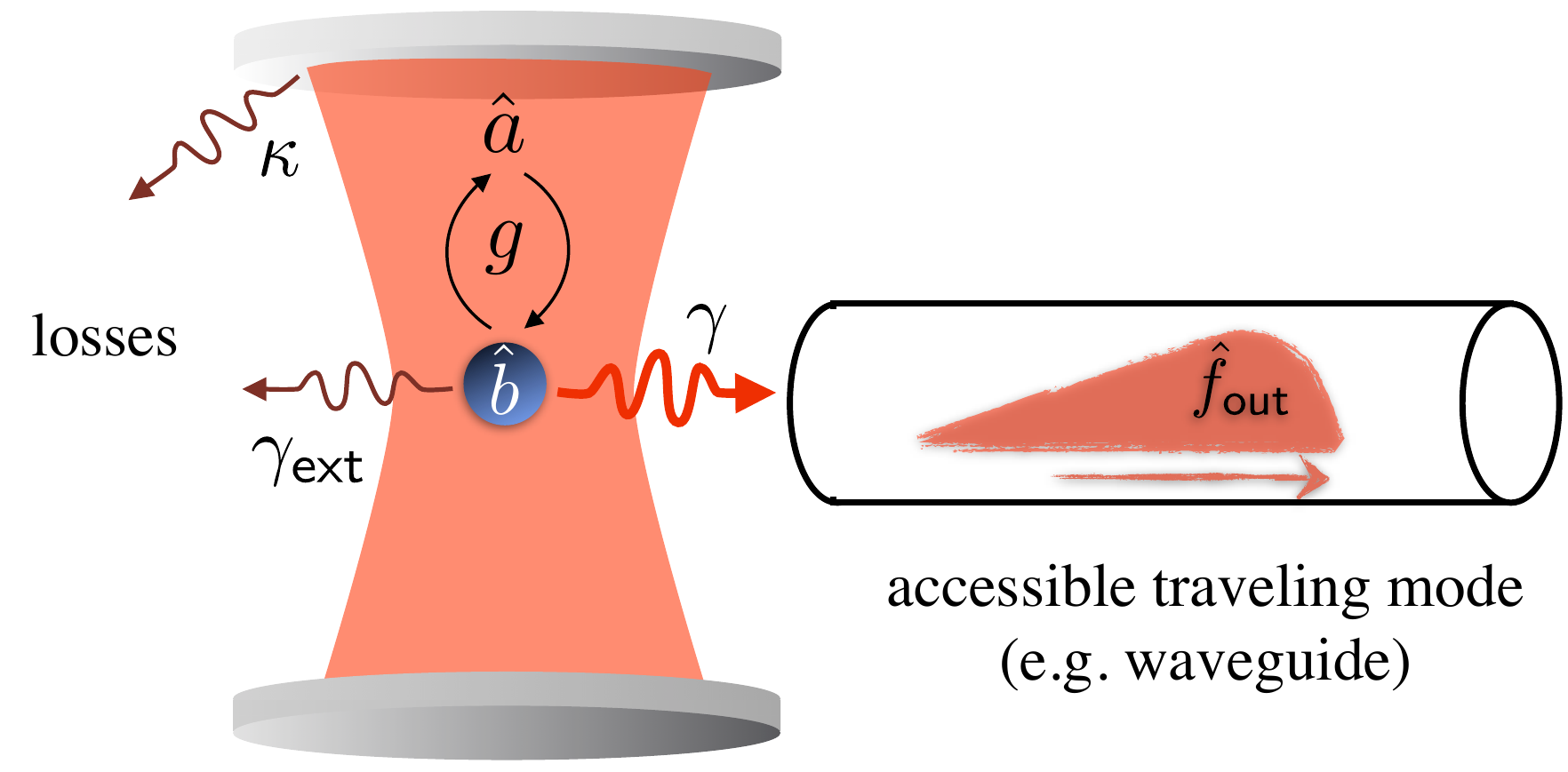}\end{center}
\caption{(Color online) Schematics of the proposed model. A high-Q cavity field (mode $\hat a$) and a bosonic mediator (mode $\hat b$) interact resonantly with strength $g$. Mode $\hat b$ radiates into the waveguide at rate $\gamma$, allowing the cavity field to be mapped onto the travelling mode $\hat f_{\sf out}$. Residual losses, not associated with detectable modes, are quantified by $\gamma_{\sf ext}$ for the bosonic mode and $\kappa$ for the cavity field.
\label{sketch}}
\end{figure}

\noindent In light of the above, it would be extremely desirable to control in time the Q-factor of a cavity, possibly switching between a cavity in the strong coupling regime and an ``open'' one in a coherent fashion. To this end, theoretical and experimental advances have been achieved in photonic crystal cavities \cite{ph_crystals}, and some degree of control at the quantum level has been very recently demonstrated in superconducting resonators \cite{catchmicro} and in optical cavities \cite{furusawa}. Let us stress that these efforts differ from usual studies on qubit networks \cite{ions}, despite the latter often require to release and catch photons between cavities. In fact the former aim at converting the field confined into a resonator -- a continuous-variable system -- to a travelling field, whereas the latter try to exchange information between confined two-dimensional systems (\textit{e.g.}, two-level atoms trapped in cavities with {\it fixed} Q-factor).

\noindent In this paper, we propose a general framework to achieve such coherent control of a resonator Q-factor, by introducing a mediating bosonic mode that scatters coherently the cavity radiation into an experimentally accessible, one-dimensional continuum of modes (waveguide for brevity). We quantify the performance of our scheme in terms of a few effective model parameters, and discuss some possible implementations based on the cavity-QED architecture. The complementary process of `feeding' an initially empty cavity through the waveguide is also studied and shown to yield the same performance \cite{supplemental}. 

\noindent\textit{The basic model} --- The system under investigation is sketched in Fig.~\ref{sketch}. We consider a high-Q resonator (cavity for brevity) in which a desired quantum field has been prepared in advance. In order to switch the quality factor, the cavity is brought to interact with a bosonic scatterer which, in turn, radiates into an accessible waveguide. In this way the initial quantum state of the cavity can be coherently transferred to a traveling mode of light, thus effectively ``opening" the cavity. To gain advantage from such a scheme, one has to be able to control the coupling between scatterer and cavity on short time scales: this may be obtained, e.g., by applying a detuning to the scatterer \cite{EITtunings}, or by controlling how much time it spends in the resonator. If these requirements are met, one can switch between a ``closed" and an ``open" cavity on demand. We describe the (single-mode) cavity field via the annihilation operator $\hat a$ -- with $[\hat a,\hat a^\dagger]=1$ -- and the bosonic scatterer by a second annihilation operator $\hat b$. Their interaction Hamiltonian, in a frame rotating at the cavity frequency $\omega$, is assumed of the form ($\hbar=1$)
\begin{align}
		H=g(\hat a^\dagger \hat{b}+\hat{b}^\dagger\hat a),\label{H}
\end{align}
with $g$ the cavity-scatterer coupling strength. The interaction between the scatterer and the continuum of waveguide modes, characterized by an emission rate $\gamma$, is conveniently dealt with in the framework of input-output theory~\cite{collett,yurke,inout}. In addition we take into account cavity losses at rate $\kappa$ and the spontaneous emission of the scatterer into inaccessible modes --- such as field modes that do not couple to the waveguide, or other internal degrees of freedom of the scatterer outside our control --- at a rate $\gamma_{\sf ext}$ [see Fig.~\ref{sketch}]. With standard assumptions \cite{collett}, one can derive the Heisenberg-Langevin equations
\begin{align}
\dot{\hat a}&=-ig\hat b-\frac{\kappa}{2}\hat a+\sqrt{\kappa}\hat a_{\sf in},\label{eq1}\\
\dot{\hat b}&=-ig\hat a-\frac{\gamma+\gamma_{\sf ext}}{2}\hat b+\sqrt{\gamma}\hat b_{\sf in}+\sqrt{\gamma_{\sf ext}}\hat b_{\sf ext, in},\label{eq2}
\end{align}
where all operators are time dependent (the time variable $t$ will be explicitly indicated only when its omission might be misleading). In particular, $\hat b_{\sf in}$ is associated with the waveguide modes under our control, while $\hat a_{\sf in},\hat b_{\sf ext, in}$ with inaccessible environments providing losses. All input modes are characterized by two-time commutators of the form $[\hat a_{\sf in}(t),\hat a_{\sf in}^\dagger(t')]=\delta(t\!-\!t')$, with analogous expressions for $\hat b_{\sf in},\hat b_{\sf ext,in}$. As our focus shall be the system's {\it output} into the waveguide, it becomes convenient to work with the {\it output operator} $\hat b_{\sf out}\!=\!\sqrt{\gamma}\hat b\!-\!\hat b_{\sf in}$. This is characterized by the same commutation rules as $\hat b_{\sf in}$ \cite{collett}, and conveniently describes the waveguide modes affected by the emission of the system ($\hat a_{\sf out},\hat b_{\sf ext,out}$ are defined by analogous equations but are not associated with detectable modes). We can thus rephrase Eqs.~\eqref{eq1} and \eqref{eq2} in terms of these output fields. For convenience, let us define $\hat {\mathbf{v}}=(\hat a,\hat b)^\intercal$, and $ \hat{\mathbf v}_{\sf out}\equiv(\sqrt{\kappa}\hat a_{\sf out},\sqrt{\gamma}\hat b_{\sf out}\!+\!\sqrt{\gamma_{\sf ext}}\hat b_{\sf ext,out})^\intercal$. The equations of motion then read \cite{collett}
\begin{align}
	\dot{\hat {\mathbf{v}}}&=\mathbf{M}\hat {\mathbf{v}}-\hat {\mathbf{v}}_{\sf out},\label{EOM2}\\
	{\bf M}&\equiv\left(\begin{array}{ccc}
	\frac{\kappa}{2}&-ig\\
	-ig& \frac{\gamma+\gamma_{\sf ext}}{2}
	\end{array}\right).\label{M}
\end{align}

\noindent{\it Opening the cavity} --- Having fixed the notation, we can tackle the problem of ``opening" the cavity as follows. At time $t\!=\!0$, we assume that the cavity field has been prepared in a quantum state of interest, while all other relevant modes are in the vacuum. Eq.~\eqref{EOM2} can be formally integrated between times $t_0$ and $t_1$ as
$\hat{\mathbf{v}}(t_1)\!=\!{\rm e}^{{\bf M}(t_1-t_0)}\hat{\mathbf{v}}(t_0)-{\rm e}^{{\bf M}t_1}\int_{t_0}^{t_1}\!{\rm d}t'\, {\rm e}^{-{\bf M}t'}\hat{\mathbf{v}}_{\sf out}(t')$. Crucially, this expression is valid also when $t_0>t_1$. Taking $t_1\!=\!0$ and $t_0\!\to\!\infty$, and using the stability condition $\lim_{\tau\to\infty}{\rm e}^{-{\bf M}\tau}\!=\!0$, one has
\begin{align}
\hat{\mathbf v}(0)=\int_0^\infty{\rm d}t\,{\rm e}^{-{\bf M} t}\hat{\mathbf v}_{\sf out}(t),
\end{align}
which relates the system operators at time $t=0$ to specific combinations of the output fields in the time interval $[0,\infty]$. Expanding the first component of this vectorial identity as $\hat a(0)=\sqrt{\gamma}\int_0^\infty{\rm d}t\,({\rm e}^{-{\bf M} t})_{1,2}\hat{b}_{\sf out}(t)+{ \text{\small\sf [inaccessible modes]}}$, one can recast it in terms of canonical bosonic operators as
\begin{align}\label{bs1}
	\hat a(0)=\sqrt{F}\hat f_{\sf out}-\sqrt{1-F}\hat h_{\sf ext},
\end{align}
where $\hat f_{\sf out}\equiv\int^{\infty}_0{\rm d}t\,u(t)\hat b_{\sf out}(t)$ is a canonical bosonic mode of temporal profile $u(t)\equiv({\rm e}^{-{\bf M}t})_{1,2}/({\int^{\infty}_0{\rm d}t'\,|({\rm e}^{-{\bf M}t'})_{1,2}|^2)^{1/2}}$ which propagates away from the system along the waveguide, while $\hat h_{\sf ext}$ is a canonical bosonic mode representing the portion of the field that has been dissipated into the inaccessible modes $\hat a_{\sf out},\hat b_{\sf ext}$ (we do not concern ourselves with the specific form of $\hat h_{\sf ext}$, its sign being chosen for later convenience). The parameter $F$, verifying $0\!\leq F\!\leq1$ by construction, is given by
\begin{align}
F=\gamma\int^{\infty}_0{\rm d}t\,|({\rm e}^{-{\bf M}t})_{1,2}|^2.\label{F}
\end{align}
We note the mapping between Eq.~\eqref{bs1} and a beam splitter \cite{barnett} of transmittivity $F$ where $\hat f_{\sf out}$ and the inaccessible mode $\hat h_{\sf ext}$ are mixed. Since all field modes except $\hat a(0)$ were initially in the vacuum, and the global evolution conserves the total excitation number \cite{collett}, it follows that at the other output of this abstract beam-splitter one must find the vacuum. That is, the relation $\sqrt{1\!-\!F}\hat f_{\sf out}+\!\!\sqrt{F}\hat h_{\sf ext}\!=\!\hat a_{\sf vac}$ must hold, with $\hat a_{\sf vac}$ a canonical bosonic mode in the vacuum state. Therefore, by inverting these relationships, one is finally able to express
\begin{align}
	\hat f_{\sf out}=\sqrt{F}\hat a(0)+\sqrt{1-F}\,\hat a_{\sf vac}.\label{fout}
\end{align}
The explicit identification of the mode $\hat f_{\sf out}$ is a crucial result since it provides by construction the traveling mode that best retains the quantum information of the cavity field $\hat a(0)$. Clearly, the larger $F$ is, the closer the output field  $\hat f_{\sf out}$ is to the initial cavity field. In addition, the Schr\"{o}dinger picture interpretation of Eq.~(\ref{fout}) is straightforward: suppose the cavity is prepared in the state $\rho_0$ at time $t\!=\!0$; the mode $\hat f_{\sf out}$ is then found in the state $\rho_{\sf out}\!=\!{\rm e}^{(1\!-\!F)\mathcal L}\rho_{0}$ where $\mathcal L\rho=\tfrac{1}{2}\left(2\hat a\rho \hat a^\dagger\!-\!\hat a^\dagger\hat a\rho\!-\!\rho\hat a^\dagger\hat a\right)$. As an instructive example, in \cite{supplemental} we apply these ideas to study the extraction of squeezed light from a cavity. To summarize, we mapped the opening of a high-Q cavity to a simple beam-splitter evolution, or equivalently, to an amplitude damping channel. These are defined by a single parameter $F$, which subsumes the details of the model, and qualifies as the natural figure of merit of our scheme. 
Defining $\gamma_{\sf tot}\!\equiv\!\gamma\!+\!\gamma_{\sf ext}$ one finds \cite{supplemental}
\begin{equation}
	F=\frac{1-\frac{\gamma_{\sf ext}}{\gamma_{\sf tot}}}{1+\frac\kappa{\gamma_{\sf tot}}+\frac{\gamma_{\sf tot}\kappa}{4g^2}+\frac{\kappa^2}{4g^2}}.\label{Fido}
\end{equation}
Notice that $F$ is monotonically decreasing in $\kappa$, and for an ideally closed cavity ($\kappa=0$) the Q-switch approaches a perfect extraction of the cavity field, provided $\gamma_{\sf ext}\ll\gamma_{\sf tot}$. In other words, ``the more a cavity is closed, the better it can be opened". More in detail, Eq.~\eqref{Fido} illustrates the constraints that the system has to satisfy in order to obtain a high figure of merit $F\sim1$. The conditions $\gamma_{\sf ext}\ll\gamma$, $\kappa\ll\gamma$, and $\kappa\ll g$ trivially state that the decay of the system into inaccessible modes should be slow, as compared to the timescales of the desired interactions $g,\gamma$. A more specific condition can be identified, which for convenience we write as $4g^2/\kappa\gamma_{\sf tot}\gg1$. Drawing an analogy with standard cavity QED, this may be interpreted as the requirement of a large {\it cooperativity parameter} for the cavity-scatterer system: despite the constructive role of the decay rate $\gamma$, we are still requiring the system to be in a form of strong coupling regime. Quite remarkably, we find that a similar performance may be obtained if the bosonic mediator is replaced by a two level system whose excited state is only virtually populated \cite{supplemental}.

\begin{figure}[t!]
\begin{center}\includegraphics[width=.385\linewidth]{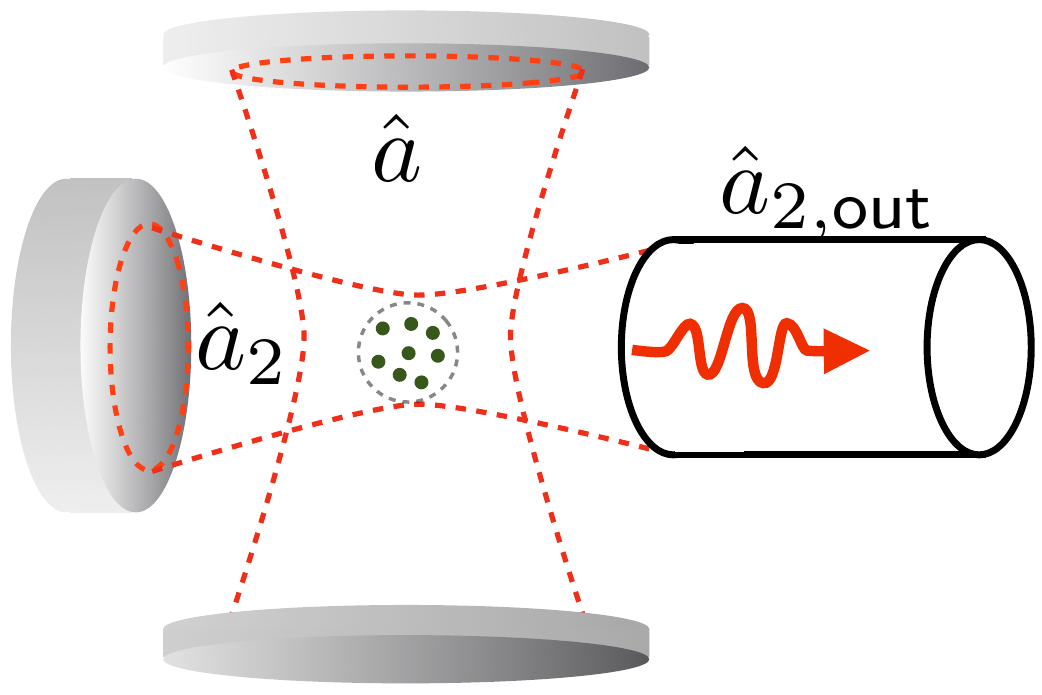}\hspace{1cm}\includegraphics[width=.385\linewidth]{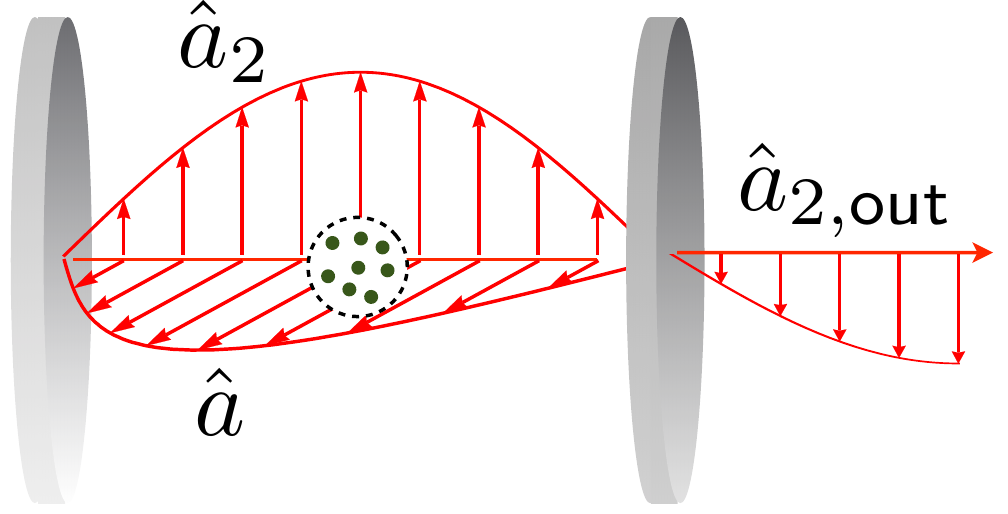}\end{center}
\caption{(Color online) An atomic ensemble interacts with two field modes of different lifetime. {\bf Left:} The fields $\hat a,\hat a_2$ belong to two different cavities. The sketch is inspired by fiber-cavity setups \cite{fibercavity}, where an optical fiber can provide both the waveguide and the transmissive mirror of the second cavity. {\bf Right:} The two modes belong to the same cavity, and are distinguished by some relevant degree of freedom such as polarization. While mode $\hat a$ is long-lived, mode $\hat a_2$ is significantly transmitted through one of the cavity mirrors. \label{figuradue}}
\end{figure}

\noindent{\it Two-mode implementations} --- Several implementations of our scheme can be envisaged, depending on the specific physical system that constitutes the high-Q resonator. We speculate here on two possible implementations based on the interaction between an ensemble of atoms and two field modes of different lifetime, as sketched in Fig.~\ref{figuradue}. As before, $\hat a$ indicates the high-Q cavity mode, while  $\hat a_2$ represents a second field mode of the same frequency. We assume that the decay of the latter is associated with emission into a waveguide at rate $\eta$, plus some optical loss at rate $\eta_{\sf ext}$. As shown in Fig.~\ref{figuradue}, the two fields may belong to different cavities, or they could be two distinct modes of the same cavity, e.g. with different polarization (in this case, the mirror transmittivity has to be different for the two modes to allow $\kappa\neq\eta$). 

{\it Setting 1.} We consider an ensemble of $n$ two-level atoms that are brought to resonance with modes $\hat a$ and $\hat a_2$ when the high-Q mode needs to be extracted. When instead a closed cavity is required, one can either apply a large detuning to the atoms or remove them altogether. We take the atoms to be initially in the ground state and identically coupled to the cavity fields, such that the interaction picture Hamiltonian reads:
\begin{align}
	H_1=\sum_{k=1}^n \left[\lambda \left( \hat \sigma^+_k \hat a + \hat \sigma^-_k \hat a^\dagger \right) +  \lambda' \left( \hat\sigma^+_k \hat a_2 + \hat\sigma^-_k \hat a_2^\dagger \right) 
	\right]\label{H1}
\end{align}
where  $\hat \sigma_k^+=(\hat \sigma_k^-)^\dagger\equiv\kebra{e_k}{g_k}$ and $\ket e_k,\ket g_k$ are the excited and ground states of the $k$-th atom, whereas $\lambda$ ($\lambda'$) denotes the coupling between the atoms and the $\hat a$ ($\hat a_2$) field. We assume here the Holstein-Primakoff regime, where $n$ is large enough, and the majority of atoms remain in the ground state during the interaction (in particular this is guaranteed when the initial cavity excitation $\langle\hat a^\dagger\hat a\rangle$ is smaller than the total number of atoms $n$), such that a collective (approximately) bosonic operator $\hat c= \frac{1}{\sqrt{n}}\sum_{k=1}^n\hat\sigma^-_k$ can be introduced (with $[\hat c, \hat c^\dagger]\simeq 1$, see \cite{hammerer10}). Denoting with $\Gamma$ the atomic decay rate into inaccessible modes, one has that the evolution of the three bosonic operators $\hat{\mathbf{v'}}\equiv(\hat a,\hat a_2,\hat c)^\intercal$ is now given by $\dot{\hat {\mathbf{v}}}'=\mathbf{M'}\hat {\mathbf{v}}'-\hat {\mathbf{v}}_{\sf out}'$ where $ \hat{\mathbf v}_{\sf out}'\equiv(\!\sqrt{\kappa}\hat a_{\sf out},\!\sqrt{\eta}\hat a_{2,\sf out}\!+\!\!\sqrt{\eta}_{\sf ext}\hat a_{2,\sf ext,out},\!\sqrt{\Gamma}\hat c_{\sf out})^\intercal$,
	\begin{align}
	{\bf M'}&\equiv\left(\begin{array}{ccc}
	\frac{\kappa}{2} & 0 & -i\lambda\sqrt n\\
	0 & \frac{\eta_{\sf tot}}{2} & -i\lambda' \sqrt{n} \\
	-i\lambda\sqrt n & -i\lambda' \sqrt{n} &  \frac{\Gamma}{2}
	\end{array}\right)\;,\label{M'}
	\end{align}
and $\eta_{\sf tot}\equiv\eta+\eta_{\sf ext}$. Our model can then be obtained by assuming the second cavity to be in the Purcell regime (or low-Q regime) \cite{porco}, namely $\eta_{\sf tot}\gg\lambda'\sqrt n$, which allows the mode $\hat a_2$ to be adiabatically eliminated. After this operation, upon identifying $\hat b\equiv\hat c$ and $\hat b_{\sf out}\equiv\hat a_{2,\sf out}$, one can finally recover Eqs. \eqref{EOM2} and \eqref{M} with $g=\sqrt n\lambda,\gamma=4n\lambda'^2\eta/\eta_{\sf tot}^2,\gamma_{\sf ext}=\Gamma+4n\lambda'^2\eta_{\sf ext}/\eta_{\sf tot}^2$. Thus, the atomic ensemble takes the role of the bosonic scatterer, while the second mode provides a means to collimate the atomic radiation into the modes of interest. Taking a step further, we find it worthwhile to study the full model described by Eq.~\eqref{M'}. Carrying out an analysis analogous to that leading to Eq.~\eqref{fout}, one arrives again at the conclusion that the process of opening the cavity can be mapped to a beam-splitter, with transmittivity $F'$ given by
\begin{align}
F'={\cal T}\eta_{\sf tot}\int^{\infty}_0{\rm d}t\,|({\rm e}^{-{\bf M'}t})_{1,2}|^2,\label{F'}
\end{align}
where ${\cal T}\equiv\eta/\eta_{\sf tot}$ is the waveguide coupling efficiency. This provides a more refined description of the process, valid beyond the Purcell regime. 
The analytical expression for $F'$ is given in \cite{supplemental}  and proves to be rather involved. Still it retains the relevant feature of approaching ${\cal T}$ for vanishing $\Gamma$ and $\kappa$.

\begin{figure}
\begin{center}
\includegraphics[width=.48\linewidth]{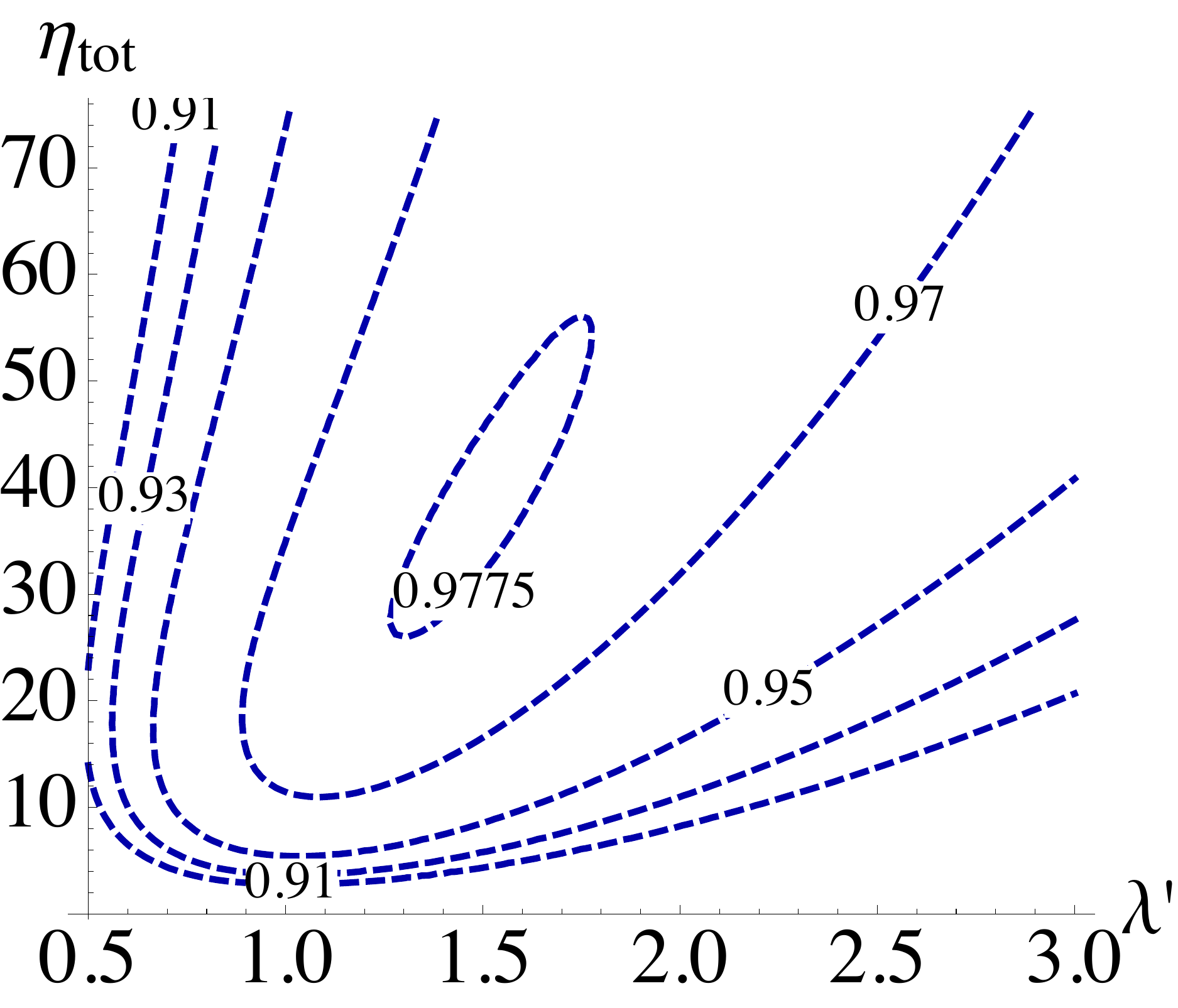}\includegraphics[width=.48\linewidth]{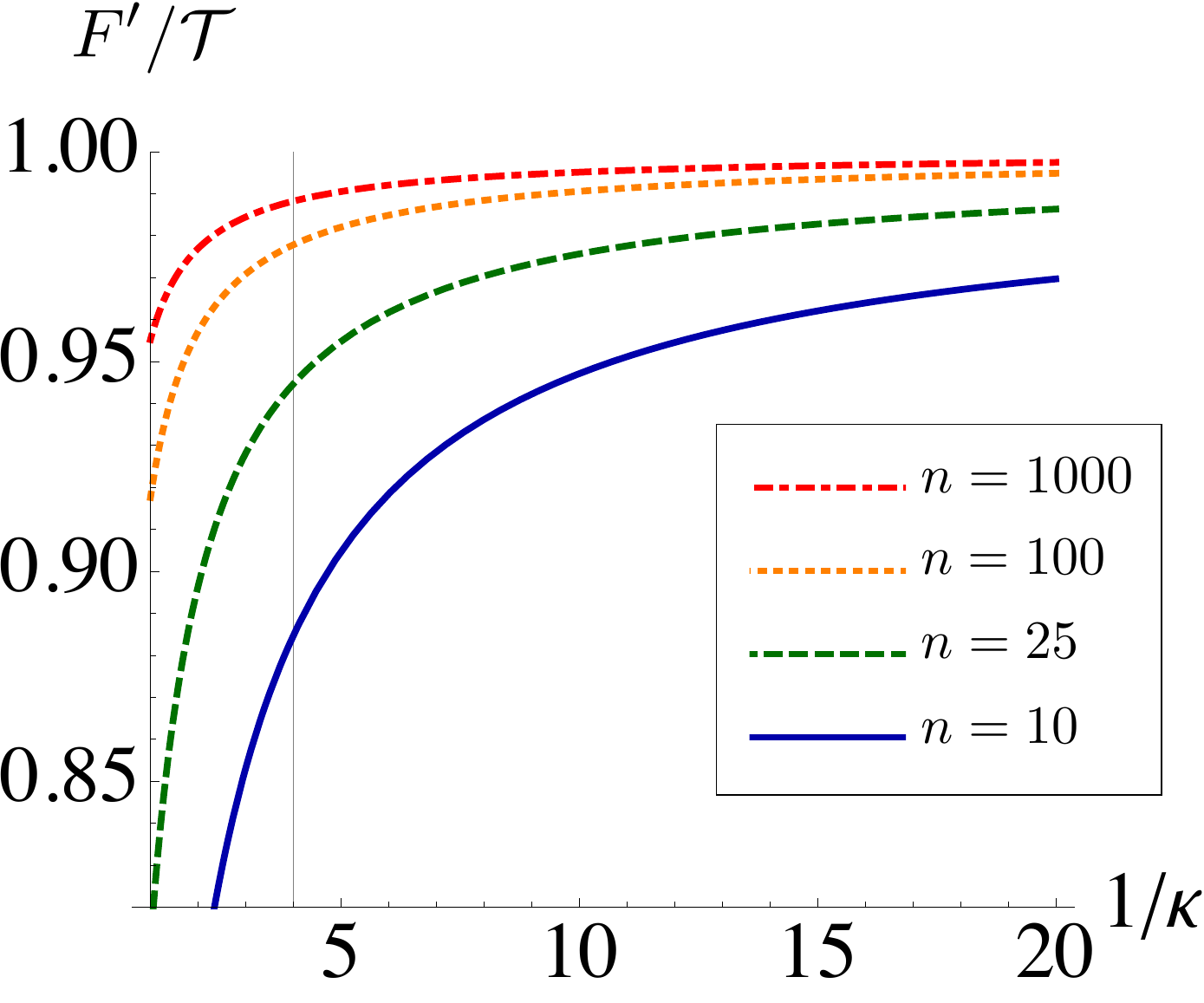}
\end{center}
\caption{(Color online). Study of the figure of merit $F'$[Eq. (\ref{F'})]. {\bf Left}: Contour plot of $F'/{\cal T}$ as a function of the parameters $\eta_{\sf tot},\lambda'$ characterizing mode $\hat a_2$. We fix units of $\lambda=1$, taking $n=100,\Gamma\simeq0.015,\kappa\simeq0.25$, as reported in Ref. \cite{BECexp}. Notice how the maximum is achieved in a region of parameters where the Purcell approximation is not accurate, $\lambda'\sqrt{n}/\eta_{\sf tot}\sim0.4$, hence our full model of Eq.~\eqref{M'} is required. {\bf Right}: Fixing $\lambda'=1.5,\eta_{\sf tot}=40$ we study $F'/{\cal T}$ as a function of $1/\kappa$, the lifetime of cavity $\hat a$, for different numbers of atoms (see inset). The vertical line indicates $\kappa\simeq 0.25$ as in Ref. \cite{BECexp}.
\label{figaura2}}
\end{figure}

In Fig.~\ref{figaura2} we report a case study inspired by the BEC-cavity system demonstrated in Ref.~\cite{BECexp}. We assume to add an auxiliary cavity $\hat a_2$ to the setup, while leaving all other parameters unchanged. Due to the properties of BECs \cite{BECs}, every atom in the ensemble experiences an {\it identical} coupling to the cavity field, so that Eq.~\eqref{H1} directly applies. In the left panel we show the behavior of $F'/{\cal T}$ as a function of the parameters $\lambda',\eta_{\sf tot}$ characterizing the auxiliary cavity. The right panel shows $F'/{\cal T}$ as a function of the number of atoms $n$ and the cavity decay rate $\kappa$, having fixed $\lambda',{\eta_{\sf tot}}$ close to their optimal values. Again, we can clearly see that $F'$ monotonically increases as $\kappa$ decreases. Due to its trivial effect on the protocol performance, the coupling efficiency ${\cal T}$ has been left implicit.

{\it Setting 2.} Hamiltonian \eqref{H} can be engineered by coupling the $n$ atoms \textit{off-resonantly} to both fields $\hat a$ and $\hat a_2$ (see e.g. Ref.~\cite{prado}). At variance with the previous case no bosonization of the atoms is required, hence the scheme may be applicable also when the number of atoms $n$ is small. We consider Eq.~\eqref{H1} with $\lambda'=\lambda$, and add a detuning $\Delta$ to all atoms, resulting in a Hamiltonian $H_2\!=\!H_1\!+\!\Delta\sum_{k=1}^n\proj{e_k}$. In the large detuning limit $\Delta\gg \lambda\sqrt{n\langle\hat a^\dagger\hat a\rangle}$ one can adiabatically eliminate the atomic excited states and derive an effective Hamiltonian for the ground state subspace \cite{james,sorensen}: $H_{\sf eff}\!=\!-n\frac{\lambda^2}{\Delta}[\hat a^\dagger \hat a\!+\!\hat a^\dagger \hat a_2\!+\!\hat a_2^\dagger \hat a\!+\!\hat a_2^\dagger \hat a_2].$ Identifying $\hat b\!\equiv\! a_2$, this provides the desired interaction Hamiltonian \eqref{H}, with $g\!=\!-n\lambda^2/\Delta$ (plus a global frequency shift which can be ignored). Thus, in this setting the bosonic scatterer is provided by the second field, with $\gamma\!\equiv\!\eta,\gamma_{\sf ext}\!\equiv\!\eta_{\sf ext}$, and one can control the strength of the coupling via the atomic detuning. The detrimental effects associated to atomic spontaneous emission can be estimated via the techniques of Ref.~\cite{sorensen}. The adiabatic elimination of the atomic excited states results in an effective decay rate $\kappa'\!\equiv\! n\Gamma ({\lambda}/{\Delta})^2$ affecting the superposition of fields $\hat a\!+\!\hat b$. This requires the modification of Eq.~\eqref{M} as per
\begin{align}
	{\bf M}&\to{\bf M}''\equiv\left(\begin{array}{ccc}
	\frac{\kappa+\kappa'}{2}&-ig+\frac{\kappa'}{2}\\
	-ig+\frac{\kappa'}{2}& \frac{\gamma_{\sf tot}+\kappa'}{2}
	\end{array}\right)\;.
\end{align}
As before, we can identify a figure of merit $F''={\cal T}\gamma_{\sf tot}\int^{\infty}_0{\rm d}t\,|({\rm e}^{-{\bf M''}t})_{1,2}|^2$ (see \cite{supplemental} for its full expression). Studying this quantity with the same parameters reported in Fig.~\ref{figaura2}, and fixing $\Delta=5\lambda\sqrt{10n}$, which guarantees the consistency of our approximations for cavity states with $\langle\hat a^\dagger \hat a\rangle\lesssim$10, we find $F''\simeq0.88{\cal T}$ for $n=1000$ and $\eta\simeq4\lambda$.

\noindent{\it Integrating standard cavity QED} --- A natural question to ask is whether the standard  protocols of cavity QED are still applicable in our two-mode scheme. Furthermore, it would be convenient if the same atoms employed for the Q-factor control could be used for this purpose. In \cite{supplemental} we show that this is indeed possible, by addressing two internal transitions of the atoms such that each mode is coupled to a different transition. By applying appropriate Stark shifts to the atoms, one is then able to control whether the atoms interact with mode $\hat a$ only, realizing cavity QED in the strong coupling regime, or with both fields $\hat a,\hat a_2$, as required for our Q-switching proposal.

\noindent{\it Conclusions and outlook} --- We have proposed a general scheme in which a mediator allows to switch coherently from a `closed' to an `open' cavity, so that the advantages of both regimes may be combined in a single setup. After having identified the accessible output mode that best represents the initially prepared cavity field, we have fully characterized the effective `transmittivity' parameter which encodes the quality of the process. As clarified in \cite{supplemental}, the same figure of merit is obtained for the complementary process of `feeding' an initially empty cavity. 
Let us also emphasise that our scheme is applicable to a single qubit mediator whose excited level is only virtually populated, which allows for an effective bosonic description \cite{supplemental}.
By considering a cavity-QED implementation we have shown that state of the art experimental parameters should be compatible with a demonstration of our scheme.

{\noindent}Our work may represent a contribution towards the achievement of ambitious goals such as the direct access to nonclassical cavity field states, the realization of cavity-based quantum memories and continuous-variable quantum networks. Goals worth pursuing include the introduction of time dependent controls to achieve time-reversal symmetry in the emitted fields, crucial for the realization of quantum networks \cite{ions}.

{\it Acknowledgments} ---  We thank G. M. Palma, F. Ciccarello, D. Segal, M.-J. Hwang, J. Hwang and M. Keller for fruitful discussions. MSK and TT acknowledge support from the NPRP 4-426 554-1-084 from Qatar National Research Fund. AF acknowledges support from the John Templeton Foundation (Grant ID 43467). AS acknowledges financial support from EPSRC through grant EP/K026267/1.

\bibliographystyle{plain}

\begin{widetext}
	\newpage
\begin{center}
\includegraphics[width=.8\textwidth]{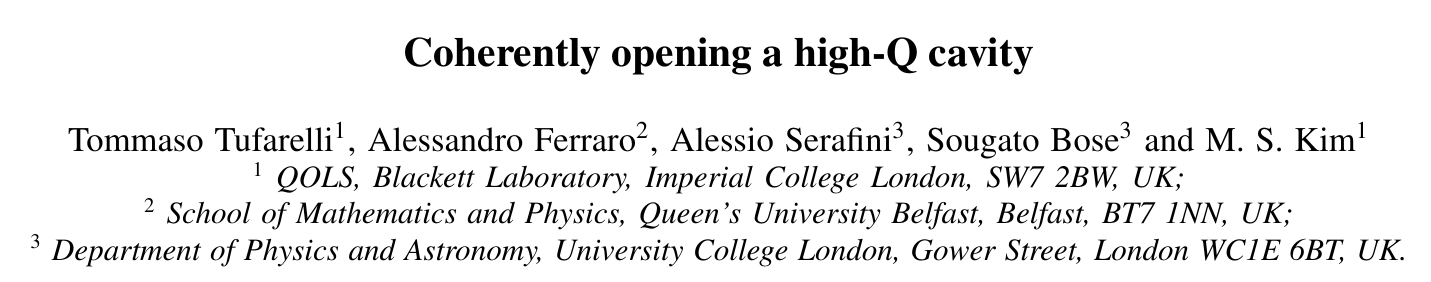}\\
\bf{SUPPLEMENTAL MATERIAL}
\end{center}
\end{widetext}
\appendix

\section{Appendix A: Feeding the cavity}
Here, we address a somewhat complementary question, as compared to the main text. We ask how well we can prepare the cavity field in a defined state, by injecting a travelling field in the waveguide. Although such a problem could be attacked by applying time-reversal arguments to the `extraction' problem studied in the main text, we prefer to give a full derivation here for clarity. The problem is conveniently formulated by expressing the equations of motion of the system in terms of input operators, as in Eqs.~\eqref{eq1} and \eqref{eq2} of the main text. For convenience, we rephrase those equation as \cite{inout}
\begin{equation}
\dot{\hat {\mathbf{v}}}={\bf A }\hat {\mathbf{v}}+\hat {\mathbf{v}}_{\sf in},\label{EOM}
\end{equation}
where
\begin{align}
{\bf A}&\equiv\left(\begin{array}{ccc}
-\frac{\kappa}{2}&-ig\\
-ig& -\frac{\gamma+\gamma_{\sf ext}}{2}
\end{array}\right),\\
\hat {\mathbf{v}}_{\sf in}&\equiv(\sqrt{\kappa}\hat a_{\sf in},\sqrt{\gamma}\hat b_{\sf in}\!+\!\sqrt{\gamma_{\sf ext}}\hat b_{\sf ext, in})^\intercal.\label{inops}
\end{align}
In this case, we assume that the waveguide field $\hat b_{\sf in}$ can be prepared in a state of choice, while $\hat a_{\sf in},\hat b_{\sf ext,in}$ are in the vacuum. In any time interval of the form $[t_0,t]$, Eq.~\eqref{EOM} can be formally integrated as
\begin{align}\label{formalint}
	\hat{\mathbf{v}}(t)={\rm e}^{A(t-t_0)}\hat{\mathbf{v}}(t_0)+{\rm e}^{At}\int_{t_0}^t{\rm d}s\, {\rm e}^{-As}\hat{\mathbf{v}}_{\sf in}(s).
\end{align}
We now consider the limits $t_0\to-\infty$ and $t\to0$ in Eq.~\eqref{formalint}. Using the stability condition $\lim_{\tau\to\infty}{\rm e}^{{\bf A}\tau}=0$, one has $\hat{\mathbf{v}}(0)=\int_{-\infty}^0{\rm d}s\, {\rm e}^{-{\bf A}s}\hat {\mathbf{v}}_{\sf in}(s)$. Considering the first component of such a vector, and recalling that the input fields $\hat a_{\sf in},\hat b_{\sf ext,in}$ are in the vacuum state, one can expand the cavity field operator at time $t=0$ as
\begin{align}\label{a0}
\hat a(0)=\sqrt\gamma\int_{-\infty}^0{\rm d}t\,\left({\rm e}^{-At}\right)_{12}\hat b_{\sf in}(t)+\text{\small{\sf [vacuum terms]}}.
\end{align}
At this point we can identify the canonical bosonic mode $\hat f_{\sf in}\equiv\int_{-\infty}^0{\rm d}t\,w(t)\hat b_{\sf in}(t),$ characterized by a temporal profile $w(t)\equiv({\rm e}^{-{\bf A}t})_{12}/({\int_{-\infty}^0{\rm d}t'\,|({\rm e}^{-{\bf A}t'})_{12}|^2)^{1/2}}$ and verifying $[\hat f_{\sf in},\hat f_{\sf in}^\dagger]=1$. This represents the particular wavepacket of waveguide modes that takes part in the determination of the prepared cavity field. On the other hand, the vacuum noise provided by the modes $\hat a_{\sf in},\hat b_{\sf ext,in}$ can be combined into a single, normalized bosonic mode $\hat a_{\sf vac}$. Hence, we have
\begin{align}\label{inputrelation}
	\hat a(0)&=\sqrt T\hat f_{\sf in}+\sqrt{1-T}\,\hat a_{\sf vac},
\end{align}
where $T\equiv \gamma\int_{-\infty}^0{\rm d}t'\,|({\rm e}^{-{\bf A}t'})_{12}|^2$. Noting that ${\bf A}=-{\bf M}^\dagger$, one can finally see that $T=F$. We thus find that also the problem of feeding the cavity can be mapped to a beam-splitter evolution, characterized by the same figure of merit that we found in the main text for the complementary task of cavity field extraction. It is easy to see that the same results apply to the models describing the two-cavity implementations presented in the main text: namely the same figures of merit $F',F''$ are found when considering the cavity feeding problem.
\section{Appendix B: Extraction of cavity squeezing}
\begin{figure}[t!]
\begin{center}
\includegraphics[width=.48\linewidth]{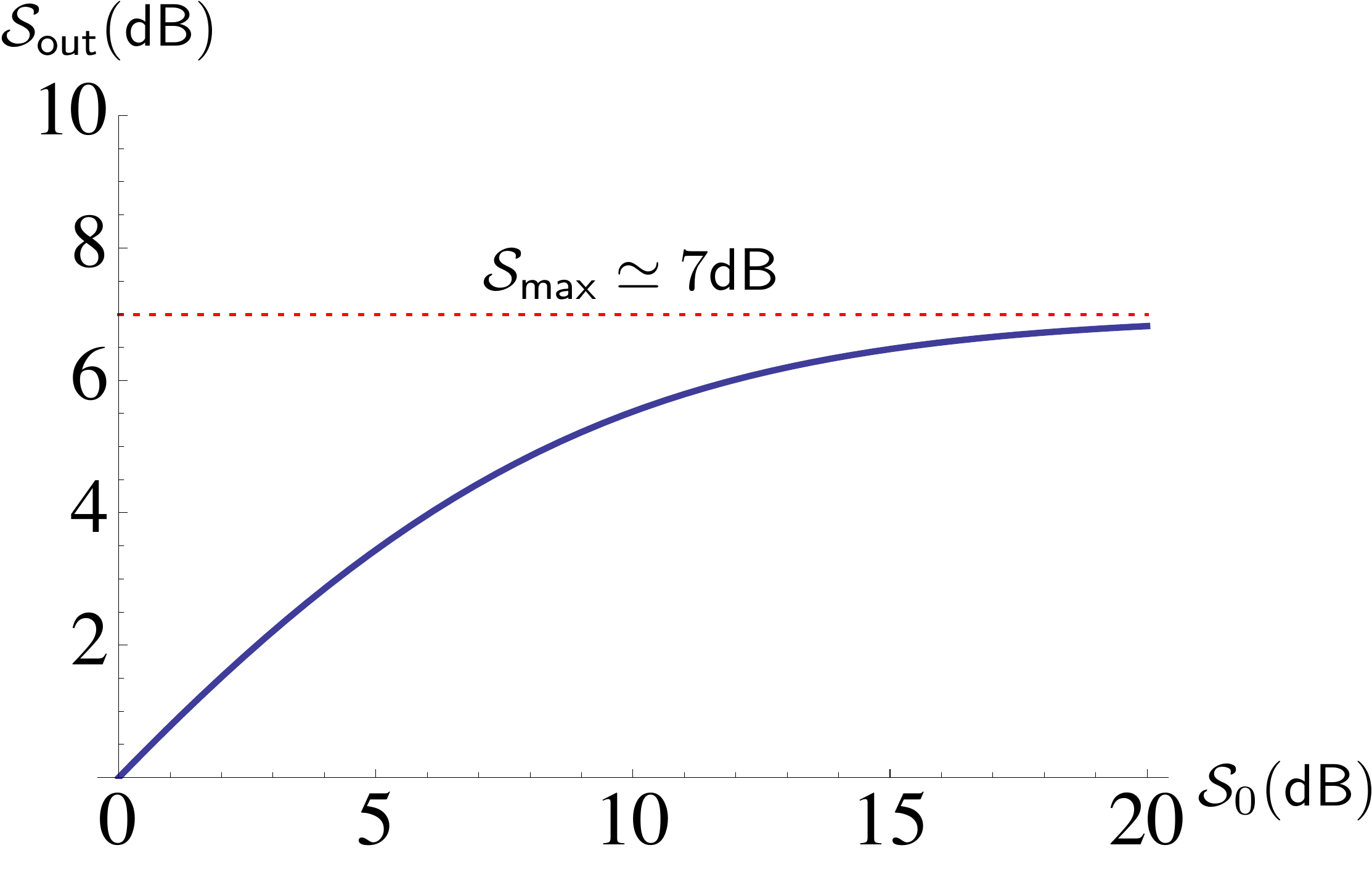}\includegraphics[width=.48\linewidth]{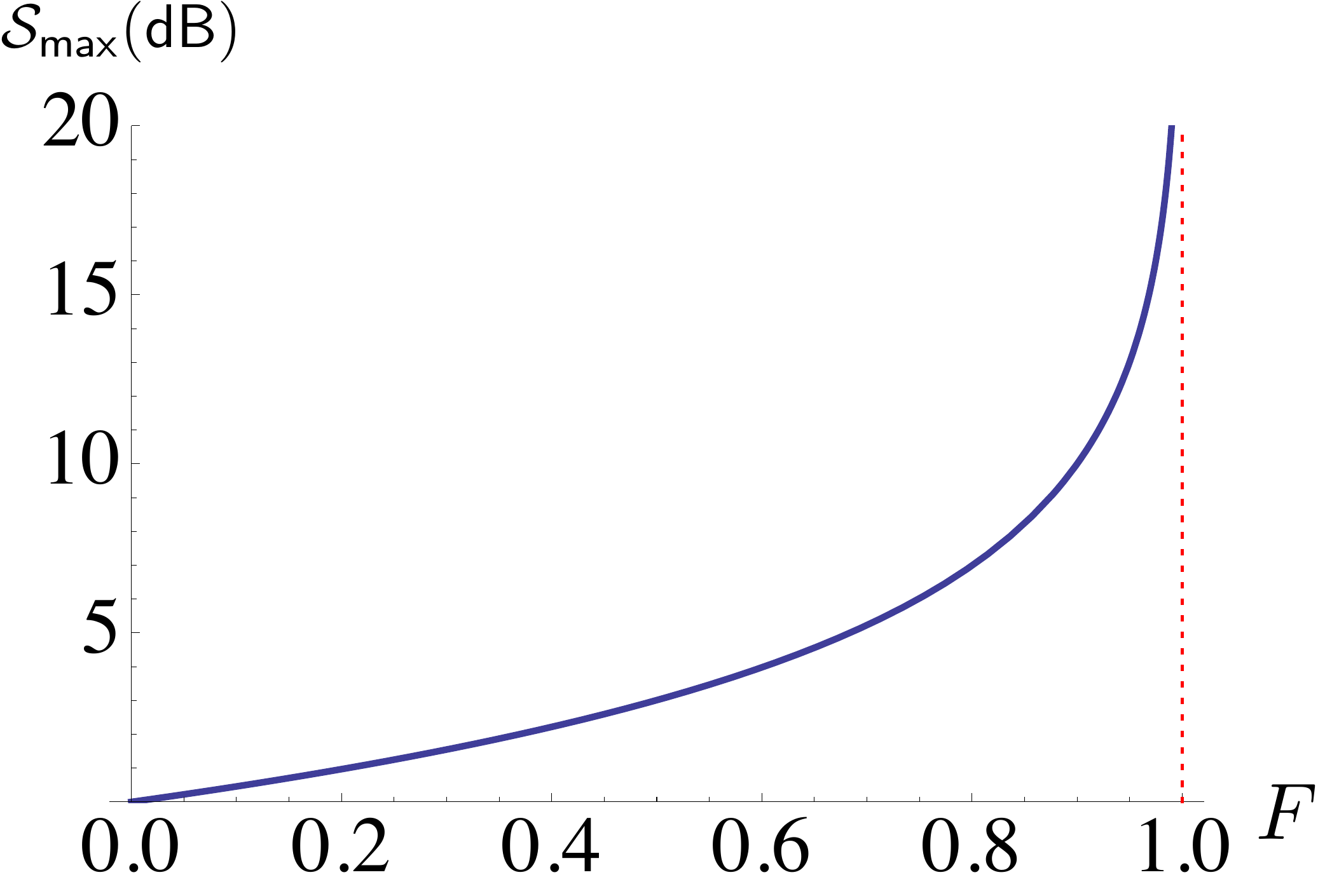}
\end{center}
\caption{(Color online) Squeezing extraction. {\bf Left:} Typical behavior of the output squeezing ${\cal S}_{\sf out}$ as a function of the initial intra-cavity squeezing ${\cal S}_0$. As ${\cal S}_0$ is increased, ${\cal S}_{\sf out}$ asymptotically reaches the value ${\cal S}_{\sf max}$ predicted by Eq.~\eqref{smax}. The figure is obtained for the particular case $F=0.8$. {\bf Right}: Upper bound to the extractable squeezing as a function of the figure of merit $F$. ${\cal S}_{\sf max}$ remains finite for any $F<1$, while it diverges for the ideal case $F\to1$ (vertical asymptote). 
\label{figsqueezzo}}
\end{figure}

As a concrete example, we shall consider the case in which the cavity mode $\hat a(0)$ is initially prepared in a squeezed state \cite{q-optics}. In particular, for such a state there is a quadrature operator of the form $\hat x_0^\theta=\hat a(0){\rm e}^{-i\theta}+\hat a(0)^\dagger{\rm e}^{i\theta}$ such that $(\Delta\hat x_0^\theta)^2<1$, where $(\Delta \hat o)^2=\langle\hat o^2\rangle-\langle\hat o\rangle^2$ is the variance. The degree of squeezing (in {\sf dB}) is defined as $\mathcal{S}(\hat x_0^\theta)\equiv-10\log_{10}(\Delta\hat x_0^\theta)^2$. Typically, in such a case one is interested in the amount of squeezing that can be transferred to the mode $\hat f_{\sf out}$. Defining the analogous quadrature operator for the waveguide mode as $\hat x_f^\theta=\hat f_{\sf out}{\rm e}^{-i\theta}+\hat f_{\sf out}^\dagger{\rm e}^{i\theta}$, and combining with Eq.~\eqref{fout} of the main text, it is easily checked that $(\Delta\hat x_f^\theta)^2=(\Delta\hat x_0^\theta)^2F+1-F$. Hence, one finds that the relationship between the initial cavity squeezing ${\cal S}_0$ and the squeezing ${\cal S}_{\sf out}$ of the output mode $\hat f_{\sf out}$ is given by
\begin{equation}
{\cal S}_{\sf out}=-10\log_{10}\left(10^{-{{\cal S}_0}/{10}}F+1-F\right).\label{sout}
\end{equation}
In particular, the above equation implies an upper bound to the extractable squeezing at fixed $F$:
\begin{equation}
{\cal S}_{\sf max}=-10\log_{10}(1-F).\label{smax}
\end{equation}
Note that, in practice, the upper bound ${\cal S}_{\sf max}$ can not be reached, as it would require the preparation of an initial cavity state with infinite squeezing. Fig.~\ref{figsqueezzo} illustrates the behavior of the output squeezing as described by Eqs.~\eqref{sout} and \eqref{smax}.
\section{Appendix C: Evaluation of the figures of merit}
We show here how the integrals entering the definition of the figures of merit $F,F',F''$ can be evaluated analytically. In general, we can study how to evaluate expressions of the form
\begin{align}
{\cal I}_{jklm}({\bf K})=\int_0^\infty {\rm d}s\,({\rm e}^{-{\bf K}s})_{jk}({\rm e}^{-{\bf K}^\dagger s})_{lm},
\end{align}
where $\bf K$ is a generic $m\times m$ matrix obeying the stability condition $\lim_{\tau\to\infty}{\rm e}^{-{\bf K}\tau}=0$. Once ${\cal I}({\bf K})$ is known, one can exploit $F={\cal I}_{1,2,2,1}({\bf M})$, and similar expressions for $F',F''$. First, we note that it is convenient to express the tensor ${\cal I}$ as a $m^2\times m^2$ matrix, according to the compact notation
\begin{align}
{\cal I}({\bf K})=\int_0^\infty {\rm d}s\,({\rm e}^{-{\bf K}s})\otimes({\rm e}^{-{\bf K}^\dagger s}),
\end{align}
where $\otimes$ indicates the Kronecker or direct matrix product. By direct inspection, one can see that $\mathcal I$ is the solution to a Sylvester equation. Indicating by $\mathbb I$ the $m\times m$ identity matrix, one has
\begin{align}
{\bf K}\otimes\mathbb I\,{\cal I}({\bf K})+{\cal I}({\bf K})\,\mathbb I\otimes{\bf K}^\dagger&=\int_0^\infty \!\!\!{\rm d}s\,\left(-\frac{{\rm d}}{{\rm d}s}\right)({\rm e}^{-{\bf K}s})\otimes({\rm e}^{-{\bf K}^\dagger s})\nonumber\\
&=\mathbb{I}\otimes\mathbb I.
\end{align}
Thus, all the required integrals can be obtained by solving linear systems of equations. Applying these techniques to the case ${\bf K}={\bf M}$, one finds Eq.~\eqref{Fido} of the main text. The figures of merit $F'$ and $F''$ are instead obtained by considering ${\bf K}={\bf M}'$ and ${\bf K}={\bf M}''$ respectively. Their full expressions read
\begin{widetext}
\begin{align}
F'&=\frac{{\cal T}\eta_{\sf tot}16 \lambda ^2 {\lambda'}^2 n^2 \left(\Gamma +\kappa +\eta _{\sf {tot}}\right)}{\left(4 \kappa  {\lambda'}^2 n+\eta _{\sf {tot}} \left(\Gamma  \kappa +4 \lambda ^2 n\right)\right) \left((\Gamma +\kappa ) \left(\Gamma  \kappa +4 \lambda ^2 n\right)+4 \Gamma  {\lambda'}^2 n+\eta _{\text{tot}} \left((\Gamma +\kappa )^2+4 {\lambda'}^2 n+(\Gamma +\kappa ) \eta _{\text{tot}}\right)\right)},\\
F''&=\frac{{\cal T}\gamma_{\sf tot}  \left(4 g^2+\text{$\kappa'$}^2\right) (\gamma_{\sf tot} +\kappa +2 \text{$\kappa'$})}{(\gamma_{\sf tot} +\kappa +2 \text{$\kappa'$})^2 (\gamma_{\sf tot}  (\kappa +\text{$\kappa'$})+\kappa  \text{$\kappa'$})+4 g^2 (\gamma_{\sf tot} +\kappa ) (\gamma_{\sf tot} +\kappa +4 \text{$\kappa'$})}.
\end{align}
\end{widetext}
Although providing an accurate description of our dynamical processes, the two expression above are rather involved and not easily interpreted. We thus find it instructive to consider relevant parameter regimes in which they simplify to a manageable form. Let us start with the quantity $F'$. In this case, it is reasonable to take the total emission rate of the auxiliary cavity $\eta_{\sf tot}$, as the dominant parameter. We then assume that the Hamiltonian coupling strengths $\lambda\sqrt{n},\lambda'\sqrt{n}$ are a first order perturbation with respect to $\eta_{\sf tot}$ [$\lambda\sqrt{n}/\eta_{\sf tot},\lambda'\sqrt{n}/\eta_{\sf tot}\sim\mathcal{O}(\epsilon)$], while the loss parameters $\Gamma,\kappa$ are of second order [$\Gamma/\eta_{\sf tot},\kappa/\eta_{\sf tot}\sim\mathcal{O}(\epsilon^2)$]. Here, $\epsilon$ represents a small parameter entering our perturbative expansion. We thus find that the leading term in the small parameter $\epsilon$ is
\begin{align}
F'=\frac{{\cal T}}{1+\frac{(\Gamma+\kappa)\eta_{\sf tot}}{4n(\lambda')^2}}+\mathcal{O}(\epsilon^2).\label{circaF}
\end{align}
Together with the obvious requirement ${\cal T}\sim 1$, signifying the necessity of a high coupling efficiency to the waveguide, we find again that a cooperativity parameter enters our figure of merit. In this case, a form of strong coupling regime in the auxiliary cavity is required. Interestingly, the first cavity decay enters such parameter as an additive contribution to the atomic decay rate. As before, we also notice that Eq.~\eqref{circaF} monotonically increases as $\kappa$ decreases.

Let us now move on to consider the quantity $F''$. In this case, proceeding in a similar manner, we may take $\gamma_{\sf tot} (=\eta_{\sf tot})$ as the dominant parameter, $g$ as a first order perturbation, and $\kappa,\kappa'$ as second order terms. We find
\begin{align}
F''=\frac{{\cal T}}{1+\frac{\gamma_{\sf tot}(\kappa+\kappa')}{4g^2}}+\mathcal{O}(\epsilon^2).
\end{align}
Once again, a cooperativity parameter directly affects the figure of merit of our protocol. In this case, the additional decay rate $\kappa'$ introduced by the atoms combines additively with the bare decay rate $\kappa$ of the main cavity.

\section{Appendix D: Opening the cavity with a two level atom}
Here, we show that even a two level system may be in principle capable of extracting a cavity field with high fidelity, in spite of its low dimensionality. Indeed, assume that the bosonic scatterer $\hat b$ of the main text is substituted by a two level atom, such that the Hamiltonian becomes $\hat H=g(\hat a\hat \sigma^++\hat a^\dagger\hat\sigma^-)$. As before, we assume that the atom decays at rate $\gamma$ in the waveguide, and $\gamma_{\sf ext}$ into external modes. The Heisenberg equations of the system in this case read \cite{barnett}
\begin{align}
\dot{\hat a}&=-\frac{\kappa}{2}\hat a-ig\hat \sigma^-+\sqrt{\kappa}\hat a_{\sf in},\label{qubiteq}\\
\dot{\hat\sigma}^-&=-\frac{\gamma+\gamma_{\sf ext}}{2}\hat\sigma^-+ig\sigma_{z}\hat a+\sigma_z(\sqrt{\gamma}\hat b_{\sf in}+\sqrt{\gamma_{\sf ext}}\hat b_{\sf ext, in}).\label{qubit}
\end{align}
Now, let us assume that all relevant cavity states satisfy $\langle\hat a^\dagger\hat a\rangle\leq \bar n$, with $\bar n$ fixed. If one works in the fast emission regime $\gamma\gg g\sqrt{\bar n}$, it is expected that the atom stays close to its ground state throughout the dynamics, and it can thus be adiabatically eliminated. Hence, we can substitute $\dot{\hat \sigma}^-\simeq0,\hat\sigma_z\simeq-1$in Eq.~\eqref{qubit}, which in turn allows to express $\hat\sigma^-$ as a function of $\hat a,\hat b_{\sf in},\hat b_{\sf ext,in}$. Subsituting this in Eq.~\eqref{qubiteq}, we have
\begin{align}
\dot{\hat a}&=-\frac{\kappa+\frac{4g^2}{\gamma+\gamma_{\sf ext}}}{2}\hat a+\sqrt{\kappa}\hat a_{\sf in}-i\frac{2g}{\gamma+\gamma_{\sf ext}}(\sqrt{\gamma}\hat b_{\sf in}+\sqrt{\gamma_{\sf ext}}\hat b_{\sf ext, in});
\end{align}
From the above equation, it is easy to identify the effective decay of mode $\hat a$ into the three channels given respectively by the cavity loss --- rate $\kappa$ ---, atomic loss --- rate $4g^2\gamma_{\sf ext}/{(\gamma+\gamma_{\sf ext})^2}$---, and waveguide emission --- rate ${4g^2\gamma}/{(\gamma+\gamma_{\sf ext})^2}$. It is then straightforward to prove that the usual beam-splitter mapping $\hat f_{out}=\sqrt{F_{\sf at}}\hat a_0+\sqrt{1-F_{\sf at}}\hat a_{\sf vac}$ can be retrieved, now characterized by a figure of merit $F_{\sf at}$ that is given by the ratio between waveguide emission and total emission:
\begin{align}
F_{\sf at}\simeq\frac{1-\frac{\gamma_{\sf ext}}{\gamma+\gamma_{\sf ext}}}{1+\frac{(\gamma+\gamma_{\sf ext})\kappa}{4g^2}}.
\end{align}
This indeed coincides with the result obtained for the bosonic scatterer in the main text, if the limit $\gamma\gg g\gg \kappa$ is taken in Eq.~\eqref{Fido}.
\section{Appendix E: Integrating standard cavity QED with three-level atoms}
For simplicity, let us describe here the case of $n=1$ atom, the extension to $n>1$ atoms being straightforward. Together with the ground state $\ket g$, we now consider two degenerate excited states $\ket{e_1}$ and $\ket{e_2}$. We assume that the transition $\ket g\leftrightarrow\ket{e_1}$ is coupled to the cavity field $\hat a$ only, while $\ket g\leftrightarrow\ket{e_2}$ is coupled to $\hat a_2$. In the two-cavity setup we have proposed, this selectivity may be due to a different spatial orientation of the atomic emission pattern of the two transitions, such that each transition is spatially matched to only one of the two cavity fields. If instead $\hat a$ and $\hat a_2$ belong to the same cavity, the selectivity may be due to the different polarization of the two modes. The interaction Hamiltonian between atoms and fields reads
\begin{align}
	H_I\!=\!\lambda_1(\hat a \ket{e_1}\bra{g}\!+\!\hat a^\dagger\kebra{g}{e_1})\!+\!\lambda_2(\hat a_2 \ket{e_2}\bra{g}\!+\!\hat a_2^\dagger\kebra{g}{e_2}),
\end{align}
where $\lambda_1$ and $\lambda_2$ are the coupling strengths. To realize standard cavity QED within the high-Q cavity $\hat a$, we have to suppress the Hamiltonian term proportional to $\lambda_2$. This can be achieved by simply applying a large detuning $\Delta_2\gg\lambda_2$ to the excited state $\ket{e_2}$, which amounts to considering $H_I\to H_I+\Delta_2\proj{e_2}$. After this operation, level $\ket{e_2}$ may be neglected, and the atom may be taken as a two level system strongly coupled to cavity $\hat a$ only.

If instead we want to realize our coherent Q-switching scheme, we require the atom to behave as a two-level system that interacts with both fields $\hat a,\hat a_2$. To this end, we consider the rotated basis $\ket{e}\equiv(\ket{e_1}+\ket{e_2})/\sqrt2,\ket{e'}\equiv(\ket{e_1}-\ket{e_2})/\sqrt2$. The interaction Hamiltonian then reads
\begin{align}
	H_I=\left(\frac{\lambda_1}{\sqrt2}\hat a\!+\!\frac{\lambda_2}{\sqrt2}\hat a_2\right)\kebra{e}{g}\!+\!\left(\frac{\lambda_1}{\sqrt2}\hat a\!-\!\frac{\lambda_2}{\sqrt2}\hat a_2\right)\kebra{e'}{g}+{\sf h.c.}
\end{align}
It is now easy to see that the addition of a large detuning $H_I\to H_I+\Delta'\proj{e'}$, with $\Delta'\gg\lambda_1,\lambda_2$, allows us to neglect level $\ket{e'}$, so that Hamiltonian \eqref{H1} can be retrieved with $\lambda\equiv\lambda_1/\sqrt2,\lambda'\equiv\lambda_2/\sqrt2$.

\end{document}